\documentclass[12pt,a4paper]{iopart}

\newcommand{\eq}[1]{\begin{equation}#1\end{equation}}

\newcommand{\ee}{\mathrm{e}}

\newcommand{\Trb}{\mathrm{Tr}_B \,}
\newcommand{\rp}{\mathrm{Re \,}}
\newcommand{\ip}{\mathrm{Im \,}}
\newcommand{\Tro}{\mathrm{Tr_{\mathrm{o}} \,}}
\newcommand{\Tre}{\mathrm{Tr_{\mathrm{e}} \,}}
\newcommand{\Treo}{\mathrm{Tr_{\mathrm{e/o}} \,}}
\newcommand{\lneg}{\mathcal{E}}
\newcommand{\eo}{\mathcal{E}_{\mathrm{o}}}

\newcommand{\bea}{\begin{eqnarray}}
\newcommand{\eea}{\end{eqnarray}}
\newcommand{\beas}{\begin{eqnarray*}}
\newcommand{\eeas}{\end{eqnarray*}}

\newcommand{\twomat}[4]{\left(\begin{array}{cc} #1 & #2 \\ #3 & #4\end{array}\right)}
\newcommand{\identity}{\ensuremath{{\rm 1 \hspace{-0.25em} l}{}}}

\usepackage{iopams}
\usepackage{setstack}
\usepackage{graphics}
\usepackage{graphicx}
\usepackage{psfrag}
\usepackage{color}
\usepackage{colordvi}
\usepackage{bbm}

\begin{document}

\title{On the partial transpose of fermionic Gaussian states}

\author{Viktor Eisler$^1$ and Zolt\'an Zimbor\'as$^{2}$}
\address{$^1$MTA-ELTE Theoretical Physics Research Group,
E\"otv\"os Lor\'and University, P\'azm\'any P\'eter s\'et\'any 1/a, H-1117 Budapest, Hungary \\
$^2$Department of Computer Science, University College London, Gower Street,
{WC1E 6BT} London, United Kingdom}

\begin{abstract}
We consider Gaussian states of fermionic systems and
study the action of the partial transposition on the density matrix.
It is shown that, with a suitable choice of basis, these states are transformed into
a linear combination of two Gaussian operators that are uniquely defined in terms of 
the covariance matrix of the original state. In case of a reflection symmetric geometry, 
this result can be used to efficiently calculate a lower bound for a well-known
entanglement measure, the logarithmic negativity. 
Furthermore, exact expressions can be derived for traces
involving integer powers of the partial transpose. The method can
also be applied to the quantum Ising chain and the results show perfect agreement with
the predictions of conformal field theory.
\end{abstract}

\maketitle

\section{Introduction}

Entanglement plays a key role in the study of quantum many-body systems \cite{Amico08,CCD09}.
Considering a pure state of a composite system, a simple measure of the entanglement between two
complementary parts is given by the von Neumann (or entanglement) entropy. Particularly interesting
is the case of pure ground states where, in great generality, an area law for the entanglement emerges
\cite{ECP10}. The most well established exceptions are one-dimensional quantum chains at criticality,
where the entanglement entropy shows a universal logarithmic scaling \cite{Vidal03} which can be
fully understood with the help of conformal field theory (CFT) \cite{CC09}. The predictions of CFT
have been confirmed on a variety of lattice models, among which a distinguished role is played by
free-particle Hamiltonians. The ground states of these systems are given by bosonic/fermionic
Gaussian states where the full entanglement spectrum is easily accessible \cite{PE09}.

The characterization of entanglement for mixed states is, however, far less obvious since,
in contrast to the pure-state scenario, there is no unique way of defining a well-behaved measure.
Among the numerous proposals for entanglement measures \cite{PV07}, a large family is based on
a convex-roof extension of the von Neumann entropy. The drawback of these
constructions is that they are essentially uncomputable already for systems of relatively small size.
A viable alternative is based on an entirely different approach, making use
of a special property of the partial transposition. Namely, the spectrum of the partial transpose
of a density matrix may contain negative eigenvalues, only if the state is entangled \cite{Peres96,H3}.
In turn, a measure called logarithmic negativity \cite{VW02} can be introduced, which quantifies how much
the partial transpose of a state fails to be positive, and can be shown to fulfill all the requirements of
an entanglement measure \cite{Plenio05}.

Although being a computable measure, the evaluation of logarithmic negativity might still pose
a significant challenge in extended quantum systems. A notable exception is the case of bosonic
systems, where the effect of partial transposition is equivalent to a partial time-reversal
of the momenta in the corresponding subsystem \cite{Simon2000}.
Furthermore, the partial transpose of bosonic Gaussian states remains to be Gaussian and,
in turn, one has a simple formula to compute the logarithmic negativity via the covariance
matrix \cite{AEPW02}. Remarkably, the analogue statement does not hold for fermionic Gaussian states.

The early studies of logarithmic negativity in lattice systems were conducted for
the harmonic oscillator chain \cite{AEPW02,FCGA08,CFGA08,AW08,Anders08,MRPR09}
using the covariance matrix technique, and for spin chain models \cite{WMB09,BSB10}
via density matrix renormalization group calculations. In contrast, exact analytical results were
found only for a few simple spin models \cite{WVB10,SKB11}.
A renewed interest in the problem was triggered recently, after a systematic approach
within CFT was introduced \cite{CCT12}. This method could be applied to
calculate the entanglement negativity for various geometries in ground \cite{CCT13,NCT15}
or thermal states \cite{CCT15, EZ14} of one-dimensional systems, as well as in
out-of-equilibrium situations \cite{EZ14,CTC14,HD14,WCR15}.

Even though the predictions of CFT can be routinely tested on harmonic chains, calculating
the logarithmic negativity in fermionic or spin systems remains to be more difficult.
Recent studies employed a tensor-network representation of the partial transposition
to calculate entanglement negativity for the transverse Ising chain \cite{CTT13}.
Alternatively, Monte Carlo techniques were applied to calculate higher moments of the partial
transpose \cite{Alba13,CABCL13}. However, even for the simplest case of fermionic Gaussian
states, a method which could compete with the computational simplicity of the bosonic case
has so far been unknown.

Here we show that, with a suitable choice of basis, the partial transpose of fermionic Gaussian
states can be cast in a particularly simple form. Namely, it can be written as the linear combination
of only two Gaussian operators, uniquely defined by corresponding covariance matrices which can be
found explicitly. Under further assumption of a reflection symmetric geometry, this construction
can be used to calculate a lower bound for the logarithmic negativity via the covariance matrix spectrum.
For critical systems, the scaling behaviour of this bound shows remarkable similarities to
that of the entanglement negativity. Furthermore, the higher moments of the partial transpose
can be exactly evaluated through simple trace formulas, providing a way to test the
universal CFT predictions on fermionic Gaussian states with minimal computational costs.

In Section~\ref{Sec:Model} we define fermionic Gaussian states and introduce the specific models
in consideration. The partial transposition transformation
for fermions is discussed in Sec.~\ref{sec:pt}, focusing on a particular choice of basis which
leads directly to our main result. Sec.~\ref{sec:ptln} is devoted to the construction
of a lower bound for the logarithmic negativity, and its numerical investigation
for the quantum Ising chain.
Trace formulas for integer powers of the partial transpose of the reduced density matrix
are presented in Sec.~\ref{sec:tr}. The paper concludes in Sec.~\ref{sec:disc} with a short
discussion of the results and their possible extensions. 
Various details of analytical calculations are included in three Appendices.

\section{Model and definitions} \label{Sec:Model}

We consider quantum systems associated to free-fermion Hamiltonians
\eq{
H = \sum_{m,n=1}^{N} \left[ A_{mn} c_m^\dag c_n +
\frac{1}{2} B_{mn} c_m^\dag c_n^{\dag} -  \frac{1}{2} B_{mn}^{*} c_m c_n \right],
\label{ham}}
where the matrices $A$ and $B$ are Hermitian and antisymmetric, respectively.
The fermionic creation/annihilation operators, $c^\dag_m$ and $c_m$,
satisfy the canonical anticommutation relations $\left\{c^\dag_m,c_n\right\}=\delta_{mn}$.
For our purposes, it will sometimes be more convenient to work with Majorana fermions defined as
\eq{
a_{2m-1}=c_m + c^\dag_m, \qquad a_{2m}=i(c_m - c^\dag_m ),
\label{mf}}
satisfying the relations $\left\{a_k,a_l\right\} =2\delta_{kl}$. In terms of  Majorana operators,
the Hamiltonian of Eq.~\eref{ham} with real $A$ and $B$
can be rewritten as $H= i \sum _{m, n=1}^{2N} T_{m,n} a_m a_n$,
where $T_{2m,2n-1}{=}{-}T_{2n-1,2m}{=}\frac{1}{4}(A_{mn}{+}B_{mn})$ and
$T_{2m,2n}{=}T_{2m-1,2n-1}{=}0$.
The product of all Majorana operators define the parity operator, 
$P=i^N \prod_{n=1}^{2N} a_n$, which plays an important role in fermionic systems.      
According to the parity superselection rule, only density matrices that 
commute with $P$ correspond to physical states \cite{BCW07,BCW09, ZZKS14}. 

The states we are going to study in this paper are the so-called \emph{Gaussian states}.
These describe the ground and Gibbs states of quadratic Hamiltonians, and 
play a prominent role in quantum information theory \cite{bravyi05,BK12, de13}.
A state $\rho$ is Gaussian if it can be written as
\eq{
\rho = \frac{1}{Z} \exp ({\sum_{k,l}  \, W_{kl} a_k a_l/4})\, ,  
\label{rhog}}
where $W$ is a purely imaginary antisymmetric matrix (with the possibility of  
$|W_{kl}| \to \infty$ allowed)
and $Z$ is the normalization factor. A Gaussian state
can also be characterized uniquely by its \emph{covariance matrix }
$\Gamma_{kl} = \langle \left[ a_k,a_l\right] \rangle /2$ via
\eq{
\tanh \frac{W}{2} = \Gamma.
\label{cov}}
Using the covariance matrix, one can express the expectation value of 
any Majorana monomial through the Wick expansion 
\eq{
\Tr(\rho  \,  a_{n_1} a_{n_2} \ldots a_{n_{2\ell}}) = \sum_{\pi} 
\mathrm{sgn} \, (\pi)
\prod_{k=1}^{\ell} \Gamma_{n_{\pi(2k-1)}, n_{\pi(2k)}},
\label{wick}}
where all the indices are different, the sum 
runs over all pairings $\pi$, and $\mathrm{sgn}\, (\pi)$ denotes the sign of $\pi$ \footnote{A pairing $\pi$ over a set $\{1,2, \ldots, 2\ell\}$
is a permutation of the $2\ell$ elements which satisfies $\pi(2k-1)<\pi(2k)$ and $\pi(2k-1)<\pi(2k+1)$ for any  $1\le k \le \ell$.  Any pairing can be decomposed into 
an $M$ number of transpositions (simple exchange of only two indices), 
and the sign of the pairing is defined as $\mathrm{sgn}\, (\pi) = (-1)^M$.}.
Let us note, that similarly to Gaussian states, one can introduce 
general \emph{Gaussian operators} which are also defined through Eq.~\eref{rhog},
however,  without requiring that the spectrum of $W$ is real. The 
Wick expansion, i.e. Eq.~\eref{wick}, holds for these operators, as well.  
 
We will also study spin chain models that are related to free-fermion Hamiltonians
 through the Jordan-Wigner transformation \cite{LSM61} 
\begin{equation}
a_{2m-1}= \prod_{k=1}^{m-1} \sigma_k^z \, \sigma_m^x ,  \quad \quad a_{2m}= \prod_{k=1}^{m-1} \sigma_k^z \, \sigma_m^y,
\end{equation}
where $\sigma_m^\alpha$ (with $\alpha=x,y,z$) denote the Pauli matrices acting on site $m$. 
The prototypical example is the transverse field Ising (TI) chain,
\eq{
H_{TI} = -\frac{1}{2} \sum_{m} \left(
\sigma_m^x \sigma_{m+1}^x + h \sigma_m^z \right),
\label{hti}}
where a chain of length $N$ with open boundary conditions is considered.
Applying the Jordan-Wigner transformation, the TI Hamiltonian \eref{hti}
takes the form of Eq. \eref{ham} with
\eq{
A_{mn} = \frac{1}{2}(\delta_{m,n-1} + \delta_{m,n+1}) - h \delta_{m,n}, \qquad
B_{mn} = \frac{1}{2}(\delta_{m,n-1} - \delta_{m,n+1}).
\label{ab}}
Such a quadratic Hamiltonian can be diagonalized by a canonical transformation,
\eq{
\eta_k = \sum_{m=1}^{N}
\frac{1}{2}\left[ \phi_{k}(m) \, a_{2m-1} - i\psi_{k}(m) \, a_{2m} \right],
\label{etak}}
and brought into the standard form
\eq{
H = \sum_{k=1}^{N} \Lambda_k \eta_k^{\dag} \eta_k+ \mathrm{const}.
}
The spectrum $\Lambda_k$ and the vectors $\phi_k, \psi_k$ in Eq.~\eref{etak}
follow from the eigenvalue equations
\begin{eqnarray}
(A-B)(A+B) \phi_k &= \Lambda_k^2 \phi_k , \label{phi}\\
(A+B)(A-B) \psi_k &= \Lambda_k^2 \psi_k . \label{psi}
\end{eqnarray}
With the full solution of the problem at hand, one can directly write down the
covariance matrix for a Gibbs state, with inverse temperature $\beta$,
of the open TI chain as
\eq{
\Gamma = \left[
\begin{array}{cccc}
\Pi_{11} & \Pi_{12} & \cdots & \Pi_{1N} \\
\Pi_{21} & \Pi_{22} & & \vdots \\
\vdots & & \ddots & \vdots \\
\Pi_{N1} & \cdots & \cdots & \Pi_{NN}
\end{array} \right], \qquad
\Pi_{mn} = \left(
\begin{array}{cc}
0 & -ig_{nm} \\
ig_{mn} & 0
\end{array} \right),
\label{gamma}}
with matrix elements given by
\eq{
g_{mn}= \sum_{k} \psi_k(m) \phi_k(n) \tanh \frac{\beta \Lambda_k}{2}.
\label{gmn}}

In our studies we will also be interested in the periodic TI chain, given by a
Hamiltonian as in Eq. \eref{hti} with the sum running up to $L$ and
boundary condition $\sigma^x_{L+1}=\sigma^x_1$. Note that,
for clear distinction, we will use $L$ instead of $N$ for the length of the
periodic chain. It is well known, that its ground state is the same as for the
fermionic model with matrices as in \eref{ab} but with antiperiodic boundary
conditions. The solutions of the system \eref{phi} and \eref{psi} are
plane waves, $\phi_k(m) \sim \ee^{ip_km}$ and $\psi_k(m) \sim \ee^{i\theta_k} \ee^{ip_k m}$,
with the Bogoliubov angles and eigenvalues given by
\eq{
\ee^{i\theta_k} = \frac{h-\ee^{ip_k}}{\Lambda_k}, \qquad
\Lambda_k = \sqrt{1+h^2-2h\cos p_k},
}
and the allowed values of the momenta are $p_k = (2k-1)\pi/L$ with $k=-L/2+1, \dots, L/2$.
One can also work directly in the thermodynamic limit, $L\to\infty$, where the momenta
become continuous and the sum in Eq.~\eref{gmn} for the matrix elements $g_{mn}$ is
replaced with an integral.

\section{Partial transpose for free fermions\label{sec:pt}}

As discussed in the Introduction, the partial transposition plays 
an important role in quantum information theory.  
In the context of entanglement theory, it was first studied for qubit
and qudit systems \cite{Peres96,H3}, but later also bosonic
\cite{Simon2000, WW01, GKLC01,W12}  and fermionic models
\cite{BCW07,BCW09, BR04, BFM14} were investigated.
An important result coming from these studies was that
the partial transpose of a bosonic Gaussian state is again a Gaussian operator; 
this simplifies the calculation of the negativity \cite{AEPW02, SAI05}. The analogous result
for fermionic Gaussian states does not hold, which can already be demonstrated by 
$2$-site systems, see \ref{app2}. 

This section is devoted to the derivation of a weaker, but still useful, result for the fermionic case. After recalling the notion of the partial transpose for spin systems and the corresponding definition for fermions in Section~\ref{subsec:pt1}, we show in Section~\ref{subsec:pt2} that  
the partial transpose of a Gaussian state (in a particular basis) can always be decomposed as 
a sum of two Gaussian operators. This decomposition lies at the heart of all the results
in the further sections.

\subsection{Definition of the partial transpose}\label{subsec:pt1}

%
\begin{figure}[thb]
\center
\includegraphics[width=0.8\columnwidth]{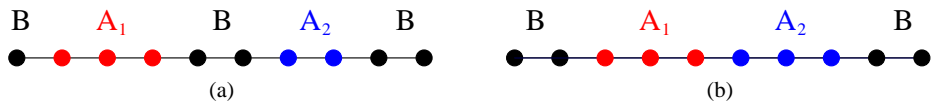}
\caption{Two possible partitioning of a spin/fermionic chain into subsystems $A_1$, $A_2$, and $B$, as described in the text.}
\label{fig:subsys}
\end{figure}
%

Consider a general  tripartition of a 
chain of qubits into disjoint 
sets $A_1$, $A_2$ and $B$, e.g. as in 
Fig.~\ref{fig:subsys}. Let $\rho$ denote the density matrix of the whole composite system.
Defining $A = A_1 \cup A_2$, the reduced density matrix (RDM) of subsystem 
$A$ is given by $\rho_A = \Trb \rho$.
The partial transpose of the RDM, $\rho_A^{T_2}$, with respect to the subsystem $A_2$ is defined
by its matrix elements as
\eq{
\langle e_i^{(1)} e_j^{(2)} | \rho_A^{T_2} | e_k^{(1)} e_l^{(2)} \rangle =
\langle e_i^{(1)} e_l^{(2)} | \rho_A | e_k^{(1)} e_j^{(2)} \rangle,
}
where $\{ | e_i^{(1)}\rangle \}$ and $\{ | e_j^{(2)}\rangle \}$ 
denote complete bases on the Hilbert spaces
$\mathcal{H}_1$ and $\mathcal{H}_2$ pertaining to the subsets $A_1$ and $A_2$.  The
definition of $\rho_A^{T_2}$ is basis dependent. However, one can easily characterize 
the set of transpositions on the operators
acting on $\mathcal{H}_2$
as those non-degenerate linear transformations that satisfy
\begin{equation}\label{eq:transposition}
\mathcal{R}(M_1 M_2) = \mathcal{R}(M_2) \mathcal{R}(M_1)\, ,  
\end{equation}
for any two operators $M_1$ and $M_2$ acting on $\mathcal{H}_2$.
Since any two partial transpositions can be connected by a unitary 
conjugation, the eigenvalues of $\rho_A^{T_2}$ are 
independent of the  choice of basis. Moreover, it was shown that the partial transpose of the
density matrix can only have negative eigenvalues if the corresponding state is 
entangled \cite{Peres96, H3}. 

In a similar way, one can define the partial transpose for fermionic states. 
Consider a tripartition of a system with $N$ fermionic modes, 
e.g. as in Fig.~\ref{fig:subsys} . Let $\{m_1, m_2, \ldots ,m_{2k} \}$ and  $\{n_1, n_2, \ldots ,n_{2\ell} \}$ denote the indices of the Majorana operators belonging to the subsystems
$A_1$ and $A_2$, respectively. Let us introduce the notation $a_x^0=\identity$ and 
$a_x^1=a_x$. A general fermionic state on $A=A_1 \cup A_2$
can be written as 
\begin{equation} \label{eq:mon_exp}
\rho_{A}= \sum_{\underline{\kappa}, \underline{\tau}} 
w_{\underline{\kappa}, \underline{\tau}} \, 
a_{m_1}^{\kappa_1}  \ldots a_{m_{2k}}^{\kappa_{2k}} a_{n_1}^{\tau_1} \ldots a_{n_{2\ell}}^{\tau_{2\ell}} \, ,
\end{equation}
where the variables 
$\underline{\kappa}=(\kappa_1, \ldots \kappa_{2k})$ and 
$\underline{\tau}=(\tau_1, \ldots , \tau_{2\ell})$ in the summation
run over all bit-strings of length $2k$ and 
$2\ell$, respectively. Note that since physical fermionic states must commute 
with the parity operator, as discussed in Section~\ref{Sec:Model}, one has
$w_{\underline{\kappa}, \underline{\tau}}$=0 when 
$\sum_{i=1}^{2k} \kappa_{i}+\sum_{j=1}^{2\ell}\tau_j$ is odd.

The partial transpose of $\rho_A$ is simply a transformation that leaves the
 $A_1$ Majorana modes invariant and acts as a transposition 
on the operators built up from modes of $A_2$, i.e.
\begin{equation}
\rho_A^{T_2}=
\sum_{\underline{\kappa}, \underline{\tau}} 
w_{\underline{\kappa}, \underline{\tau}} \, 
a_{m_1}^{\kappa_1}  \ldots a_{m_{2k}}^{\kappa_{2k}}
\mathcal{R}(a_{n_1}^{\tau_1} \ldots a_{n_{2\ell}}^{\tau_{2\ell}}) \, ,
\end{equation}
where $\mathcal{R}$ satisfies Eq.~\eref{eq:transposition}. Since also in the fermionic case 
all the transpositions are connected by a unitary conjugation, the eigenvalues 
of $\rho_A^{T_2}$ will be independent of which $\mathcal{R}$ we choose.
It will be useful to consider the following particular transposition which is defined by 
\begin{equation}
\fl
\mathcal{R}(a_{n_1}^{\tau_1} \ldots a_{n_{2\ell}}^{\tau_{2\ell}})=
 (-1)^{f(\underline{\tau})}a_{n_1}^{\tau_1} \ldots a_{n_{2\ell}}^{\tau_{2\ell}}, \; \; 
\mathrm{where} \;\;
f(\underline{\tau})=
\cases{
0 & if $|\underline{\tau}| \; \mathrm{mod} \; 4 \in \{0,3\}$,\\
1 & if $|\underline{\tau}| \; \mathrm{mod} \; 4 \in \{1,2\}$,
}
\label{ferm_part_2}
\end{equation}
where $|\underline{\tau} |= \sum_{i=1}^{2\ell} \tau_i$.
In other words, a Majorana monomial is mapped by $\mathcal{R}$ to itself if  it is of length $4n$ or $4n+3$,
and otherwise it is multiplied by a $-1$ sign. Note that a very similar transposition
in fermionic systems was already considered in Ref.~\cite{SL14}.  
Although the definition of entanglement in fermionic systems
is somewhat different from the case of spin systems, it has been proven that 
$\rho_A^{T_2}$ can only have negative eigenvalues if $\rho_A$ is entangled 
also in the fermionic case \cite{BCW07, BCW09}.

Finally, let us shortly discuss the connection between reduced density matrices of
fermionic and spin models that are connected by the Jordan-Wigner transformation.
As this transformation is non-local, it has been shown that 
the reduced density matrices corresponding to a region $A_1 \cup A_2$ 
in a spin chain model  and its fermionic counterpart are usually not equivalent
(not isospectral), unless $A_1$ and $A_2$ are adjacent intervals, as depicted in Fig.~\ref{fig:subsys}(b) \cite{FC10, IP10}. 
Since the same holds also for the 
transposed density matrices, when treating spin models we
will only consider the adjacent interval geometry. For the case of fermionic systems,
our results are valid for arbitrary geometries.
\\

\subsection{The Gaussian case} \label{subsec:pt2}

Consider a Gaussian state $\rho_A$ on the system  $A = A_1 \cup A_2$ 
with a covariance matrix 
\begin{eqnarray}
\Gamma_A = \twomat{ \Gamma^{11}}{\Gamma^{12}}{\Gamma^{21}}{\Gamma^{22}} \, ,
\end{eqnarray}
where $\Gamma^{11}$ and $\Gamma^{22}$ denote the reduced covariance matrices of subsystems
$A_1$ and $A_2$, respectively; while $\Gamma^{12}$ and $\Gamma^{21}$ contain 
the expectation values of mixed quadratic terms. Let $P_2$ be the parity operator 
 on subsystem $A_2$, and define the operators
$\rho_{+}=\case{1}{2}(\rho_A + P_2 \rho_A P_2)$ and 
$\rho_{-}=\case{1}{2}(\rho_A - P_2 \rho_A P_2)$. By definition, we have
that $\rho_A = \rho_{+} + \rho_{-}$.
Using the notation of Section~\ref{subsec:pt1}, 
$\rho_+$ and $\rho_-$ can be expanded as
\begin{equation}
\eqalign{
\rho_+= \sum_{\substack{\underline{\kappa}, \underline{\tau} \cr |\underline{\tau}| \; \mathrm{even}}} 
w_{\underline{\kappa}, \underline{\tau}} \, 
a_{m_1}^{\kappa_1}  \ldots a_{m_{2k}}^{\kappa_{2k}} a_{n_1}^{\tau_1} \ldots a_{n_{2\ell}}^{\tau_{2\ell}} \, , \\
\rho_-= \sum_{\substack{\underline{\kappa}, \underline{\tau} \cr |\underline{\tau}| \; \mathrm{odd}}} 
w_{\underline{\kappa}, \underline{\tau}} \, 
a_{m_1}^{\kappa_1}  \ldots a_{m_{2k}}^{\kappa_{2k}} a_{n_1}^{\tau_1} \ldots a_{n_{2\ell}}^{\tau_{2\ell}} \, ,}
\end{equation}
where the coefficients $w_{\underline{\kappa}, \underline{\tau}}$ can be obtained from $\Gamma_A$
using the Wick rule, Eq.~\eref{wick}. By linearity of the partial transpose,   $\rho_A^{T_2}=\rho_+^{T_2}
+ \rho_-^{T_2}$ follows, 
and $\rho_\pm^{T_2}$ can be obtained using Eq.~\eref{ferm_part_2}:
\begin{equation}
\eqalign{
\rho_+^{T_2}= \sum_{\substack{\underline{\kappa}, \underline{\tau} \cr |\underline{\tau}| \; \mathrm{even}}} 
(-1)^{|\underline{\tau}|/2} w_{\underline{\kappa}, \underline{\tau}} \, 
a_{m_1}^{\kappa_1}  \ldots a_{m_{2k}}^{\kappa_{2k}} a_{n_1}^{\tau_1} \ldots a_{n_{2\ell}}^{\tau_{2\ell}} \, , \\
\rho_{-}^{T_2}=\sum_{\substack{\underline{\kappa}, \underline{\tau} \cr |\underline{\tau}| \; \mathrm{odd}}} 
(-1)^{(|\underline{\tau}|-1)/2}w_{\underline{\kappa}, \underline{\tau}} \, 
a_{m_1}^{\kappa_1}  \ldots a_{m_{2k}}^{\kappa_{2k}} a_{n_1}^{\tau_1} \ldots a_{n_{2\ell}}^{\tau_{2\ell}} \, . }
\label{pm_part_trans}
\end{equation}
Let us introduce the generalized Gaussian operators $O_+$ and $O_-$,
with covariance matrices
\begin{eqnarray}
\Gamma_{+}=\twomat{ \Gamma^{11}}{i\Gamma^{12}}{i\Gamma^{21}}{-\Gamma^{22}}  \, , \qquad
\Gamma_{-}=\twomat{ \Gamma^{11}}{-i\Gamma^{12}}{-i\Gamma^{21}}{-\Gamma^{22}}  \, ,
\label{gammapm}
\end{eqnarray}
and consider the Majorana monomial expansion 
of these operators,
\begin{eqnarray}
 O_{\pm}=\sum_{\underline{\kappa}, \underline{\tau}} 
o^{\pm}_{\underline{\kappa}, \underline{\tau}} \, 
a_{m_1}^{\kappa_1}  \ldots a_{m_{2k}}^{\kappa_{2k}} a_{n_1}^{\tau_1} \ldots a_{n_{2\ell}}^{\tau_{2\ell}} \, .
\end{eqnarray}
Since $O_+$ and $O_-$ are Gaussian operators, one can again obtain  $o^{\pm}_{\underline{\kappa}, \underline{\tau}}$ from $\Gamma_{\pm}$ using 
Eq.~\eref{wick}. Connecting the Wick-expansion form of $w_{\underline{\kappa}, \underline{\tau}}$  
with that of $o^{\pm}_{\underline{\kappa}, \underline{\tau}}$, using the relation
between $\Gamma_A$ and $\Gamma_{\pm}$, one can deduce that
\begin{equation}
o^{\pm}_{\underline{\kappa}, \underline{\tau}}=
\cases{
\pm i (-1)^{(|\underline{\tau}|-1)/2} w_{\underline{\kappa},\underline{\tau}}  & when $|\underline{\tau}|$ odd, \\
\phantom{\pm i}(-1)^{|\underline{\tau}|/2}w_{\underline{\kappa}, \underline{\tau}} & when $|\underline{\tau}|$ even.
}
\end{equation}
Comparing this with Eq.~\eref{pm_part_trans} it immediately follows that
$\rho^{T_2}_{+}= \frac{1}{2} (O_+ + O_-)$ and $\rho^{T_2}_{-}= \frac{i}{2} (O_- - O_+)$.
Thus, we obtain the decomposition 
\begin{equation}
\rho^{T_2}_A= \frac{1-i}{2} O_+ + \frac{1+i}{2} O_- \, ,
\label{ptrho}
\end{equation}
which is, from a conceptual point of view, the main result of the paper.

\section{Partial transpose and logarithmic negativity
\label{sec:ptln}}

In the previous section, we have shown that the partial transpose of a Gaussian RDM can 
be written as a linear combination of only two Gaussian operators, which is the simplest 
possible form for a non-Gaussian operator.
However, since $O_+$ and $O_-$ do not commute in general, Eq. \eref{ptrho} can not
be rewritten for the eigenvalues, and thus one does not have a direct access
to the full spectrum of $\rho^{T_2}_A$. Nevertheless, there are a number of important
properties which can be deduced by a direct investigation of the covariance matrices
$\Gamma_\pm$. A particularly interesting quantity that we will study
is the \emph{logarithmic negativity} \cite{VW02}, which can be used as a measure of entanglement.

In the following we thoroughly investigate three special cases.
First, we consider the partial transpose for bipartite pure states which, although 
the results being well-known, turns out to be very instructive in understanding 
the implications of the decomposition in Eq.~\eref{ptrho}.
We proceed with the study of thermal mixed states in a reflection symmetric bipartite
geometry, which allows us to define and calculate a lower bound for the logarithmic negativity.
Finally, we report our findings for a genuine tripartite geometry. In each of the following
subsections, the validity of formula \eref{ptrho} was checked against exact numerical
calculations for the TI chain, Eq.~\eref{hti}, with a small number of spins.

\subsection{Bipartite pure states}

We first consider the simplest case with $B=\emptyset$ and a pure state on 
$A= A_1 \cup A_2$ given by $\rho=|\phi \rangle \langle \phi |$.
Since for any pure Gaussian fermionic state $\Gamma^2=\identity$ is satisfied,
it follows that $\left[\Gamma_{+},\Gamma_{-}\right]$=0, which implies that
$O_+$ and $O_-$ commute as well. Furthermore, since their eigenvalues are connected
by a complex conjugation, the spectrum of $\rho^{T_2}$ is simply given as the sum of
the real and imaginary parts of the $O_+$ eigenvalues.
Now, since $O_+$ is also Gaussian, its spectrum is uniquely determined through the eigenvalues
of $\Gamma_+$. To obtain them, we first assume without loss of generality that $|A_1| \le |A_2|$.
Furthermore, for  notational simplicity we also assume that $|A|$ is even, however, the results for the odd case 
follow trivially. Let $\pm  \mu_k$ denote the eigenvalues of the reduced covariance matrix
$\Gamma^{11}$ with $k=1, \ldots, |A_1|$ and $\mu_k \ge 0$. As shown in \ref{app1},
the eigenvalues $\pm \nu^{\pm}_k$ of $\Gamma_+$ can then be given as
\begin{eqnarray}
\nu^{\pm}_{k} =
\cases{
\mu_k \pm i \sqrt{1 - \mu_k^2} & when $k=1,\ldots, |A_1|$,\\
1 & when $k=|A_1| {+}1, \ldots, |A|/2$.
}
\label{nupm}
\end{eqnarray}
The canonical diagonalised form of the Gaussian operator $O_+$ reads
\eq{
O_+ = \prod_{k=1}^{|A|/2} \prod_{\sigma_k=\pm}
\frac{\identity + i\nu^{\sigma_k}_k b^{\sigma_k}_{2k-1}b^{\sigma_k}_{k}}{2}  ,
\label{o+diag}}
where $b^{\pm}_{j}$ are Majorana operators obtained from $a_j$ via the orthogonal transformation
which diagonalizes $\Gamma_+$. The eigenvalues of $O_+$ can be obtained according to the following rules.
First, we shall consider the various combinations of the conjugate eigenvalue pairs for each $k=1,\ldots, |A_1|$ as
\eq{
\omega^{\sigma_k \sigma'_k}_k = \frac{1+\sigma_k\nu^+_k}{2}\frac{1+\sigma'_k\nu^-_k}{2},
\label{omssk}}
with $\sigma_k =\pm$ and $\sigma'_k =\pm$. Using Eq. \eref{nupm} this yields
\eq{
\omega^{++}_k = \frac{1+\mu_k}{2}, \quad
\omega^{--}_k = \frac{1-\mu_k}{2}, \quad
\omega^{+-}_k = -\omega^{-+}_k = \frac{i}{2}\sqrt{1-\mu^2_k},
\label{omk}}
where we used the property $\nu^+_k \nu^-_k=1$.
The nonzero eigenvalues of $O_+$ can then be written down as
\eq{
\Omega_{\underline{\sigma} \, \underline{\sigma}'} =
\prod_{k=1}^{|A_1|} \omega^{\sigma_k \sigma'_k}_{k} =
\prod_{\sigma_k = \sigma'_k} \frac{1+ \sigma_k\mu_k}{2}
\prod_{\sigma_k =-\sigma'_k} \frac{\sigma_k i\sqrt{1-\mu^2_k}}{2},
\label{omss}}
where $\underline{\sigma}$ and $\underline{\sigma}'$ are the signature arrays 
corresponding to the eigenvalue. Note that the additional $\nu^{\pm}_k=1$ in Eq. \eref{nupm}
lead to further eigenvalues of $O_+$ that are all equal to zero.

The products in Eq. \eref{omss} are either real or purely imaginary and the
eigenvalues of $\rho^{T_2}$ thus follow as $\rp \Omega_{\underline{\sigma} \, \underline{\sigma}'}$
or $\ip \Omega_{\underline{\sigma} \, \underline{\sigma}'}$, respectively.
It is instructive, however, to derive the same spectrum using the Schmidt decomposition of $|\phi\rangle$.
Dividing the Hilbert space $\mathcal{H}=\mathcal{H}_1 \otimes \mathcal{H}_2$ into two parts
\footnote{When considering a tensor product structure, we always refer to 
a partition of the spin chain system constructed through the Jordan-Wigner transformation.}, one has
\eq{
|\phi\rangle = \sum_{i} \sqrt{\lambda_i} |\phi^1_i \rangle |\phi^2_i \rangle,
}
with the RDM eigenvalues $\lambda_i$ of subsystem $A_1$.
Clearly, the state is supported on a smaller Hilbert space  $\mathcal{H}_1\otimes \mathcal{H}_1$
and is invariant under the action of a flip operation defined by 
$|\phi^1_i \rangle |\phi^2_j \rangle \to  |\phi^1_j \rangle |\phi^2_i \rangle $.
The partial transpose of $\rho$,
\eq{
\rho^{T_2} = (|\phi\rangle \langle \phi |)^{T_2} =
\sum_{i,j} \sqrt{\lambda_i \lambda_j} |\phi^1_i \rangle \langle \phi^1_j| \otimes
|\phi^2_j \rangle \langle \phi^2_i|,
}
commutes also with the flip operator. Furthermore, it is easy to check that the eigenvalues and vectors are
\eq{
\rho^{T_2} |\phi^1_i \rangle |\phi^2_i \rangle = \lambda_i |\phi^1_i \rangle |\phi^2_i \rangle, \qquad
\rho^{T_2} |\phi^\pm_{ij} \rangle = \pm \sqrt{\lambda_i \lambda_j} |\phi^\pm_{ij} \rangle,
}
where we introduced the notation
\eq{
|\phi^\pm_{ij} \rangle =
\frac{1}{\sqrt{2}} (|\phi^1_i \rangle |\phi^2_j \rangle \pm |\phi^1_j \rangle |\phi^2_i \rangle)
\qquad i \ne j.
}
Note that all the positive (negative) eigenvalues correspond to even (odd) eigenvectors with respect to the flip operation.
Moreover, since $\rho$ is Gaussian, one can immediately write down the products
of eigenvalues as
\eq{
\sqrt{\lambda_{\underline{\sigma}} \lambda_{\underline{\sigma}'}}=
\prod_{\sigma_k =\sigma'_k} \frac{1+ \sigma_k\mu_k}{2}
\prod_{\sigma_k =-\sigma'_k} \frac{\sqrt{1-\mu^2_k}}{2},
\label{lsls}}
where the signature $\underline{\sigma}$ has again components $\sigma_k =\pm$.
Comparing Eq.~\eref{lsls} to Eq.~\eref{omss}, one indeed recognizes 
$\Omega_{\underline{\sigma} \, \underline{\sigma}'}$ up to the factors of $i$.

Owing  to the simple product structure of the eigenvalues in Eq.~\eref{lsls} and,
in particular, to the fact that all the negative eigenvalues are located in the odd subspace, 
one has
\eq{\Tr |\rho^{T_2}| = 1-2\Tro \rho^{T_2} =
\prod_{k}\left(1+\sqrt{1-\mu^2_k}\right),
}
where we introduced the notation $\Treo$ for traces taken over the even/odd subspace.
Thus, the pure-state logarithmic negativity is given by the simple formula
\eq{
\lneg = \ln \Tr |\rho^{T_2}| = \sum_{k} \ln \left(1+\sqrt{1-\mu^2_k}\right).
}
Considering also the expression of the R\'enyi entropy for fermionic Gaussian states,
\eq{
S_{n} = \frac{1}{1-n} \sum_{k}
\ln \left[ \left(\frac{1+\mu_k}{2}\right)^{n} + \; \left(\frac{1-\mu_k}{2}\right)^{n}\right],
\label{renyi}}
the well-known equality $\lneg=S_{1/2}$ for pure states can be confirmed directly.

\subsection{Thermal states in a bipartite geometry
\label{sec:bith}}

The simplicity of the pure-state scenario relies essentially on the property of
the Schmidt decomposition, which is automatically symmetric under the flip 
operation defined previously. Due to this, the partial transposed
state is block diagonal wrt the splitting into even and odd subspaces. 
Such a structure is missing for general bipartite mixed states, unless 
the system has a flip-type symmetry a priori. In this respect, a natural 
scenario would be to consider intervals of equal length 
$|A_1|=|A_2|=N/2$ and states that  are reflection symmetric. 
To analyse such a situation, we shall consider
Gibbs states of the open TI chain, Eq. (\ref{hti}), with a symmetric bipartitioning,
these being the simplest mixed Gaussian states where we hope to get further insight
into the structure of the partial transpose.

Considering a covariance matrix of the form \eref{gamma},
the spectrum of $\Gamma_+$ is invariant with respect 
to a sign change and complex conjugation, hence the eigenvalues can be collected into  
two families of quadruplets
\eq{
\left\{ z_k, z^{*}_k, -z_k, -z^{*}_k\right\}, \quad k\in \mathrm{(I)} \qquad
\left\{ iu_k, -iv_k, -iu_k, iv_k \right\}, \quad k \in \mathrm{(II)}
\label{quad}}
where in family (I) we choose $\rp z_k >0$ and $\ip z_k >0$, whereas $u_k > v_k$
in family (II). Note that the eigenvalues in the second family are purely imaginary
and thus their complex conjugate are automatically contained in the spectrum
of a skew-symmetric matrix, hence $u_k \ne v_k$. Although one could, in general,
have an arbitrary number of type (II) quadruplets, from the numerics we observe that
in the Ising case they are either absent or a single one appears. Moreover, this only
happens in the symmetry-broken phase, i.e., when $h<1$ in Eq.~\eref{hti}.

Analogously to the pure case in Eq. \eref{omssk}, we first assign the factors
\eq{
\eqalign{
\omega^{\sigma_k \sigma'_k}_k =
\cases{
\frac{1 + |z_k|^2 +\sigma_k 2 \rp z_k} {4}, & $\sigma_k = \sigma'_k$ \\
\frac{1 - |z_k|^2 +\sigma_k 2i \ip z_k }{4}, & $\sigma_k = -\sigma'_k$
}, \quad k \in \mathrm{(I)} \\
\omega^{\sigma_k \sigma'_k}_k =
\cases{
\frac{1 + u_k v_k +\sigma_k i(u-v) }{4}, & $\sigma_k = \sigma'_k$ \\
\frac{1 - u_k v_k +\sigma_k i(u+v) }{4}, & $\sigma_k = -\sigma'_k$
}, \quad k \in \mathrm{(II)}}
}
within each quadruplet, and the eigenvalues $\Omega_{\underline{\sigma}\,\underline{\sigma}'}$ of $O_+$
are again given in the factorized form of Eq.~\eref{omss}. Although the spectrum of
the operator $O_-$ is identical to that of $O_+$, they do not commute in general and
thus one has no direct access to the eigenvalues of $\rho^{T_2}$.
Nevertheless, the information about the even/odd parity of the eigenvectors is retained.
In fact, the reflection operator $R$, which defines the even/odd subspaces in our case,
acts on the spin operators as
$R\sigma_n^{\alpha}R^\dagger = \sigma_{N-n}^{\alpha}$ (with $\alpha=x,y,z$), 
implying the action $Rc_n^{\dagger}R^\dagger=Pc_{N-n}^{\dagger}$ on the creation operators,
where $P$ is the parity operator. 
Using this, it follows that the sign factor associated to an eigenvector of $O_+$ reads
\eq{
S_{\underline{\sigma}\,\underline{\sigma}'} = 
\rp \prod_{k} s_{\sigma_k\sigma'_k} +
\ip \prod_{k} s_{\sigma_k\sigma'_k}, \qquad
s_{\sigma_k\sigma'_k} = 
\cases{
1, & if $\sigma_k= \sigma'_k$,  \\
\sigma_k i, & if $\sigma_k= -\sigma'_k$,
}
}
which can also be verified by considering the pure state limit. The parity of the
$O_-$ eigenvectors are simply obtained through the factors $s^*_{\sigma_k\sigma'_k}$.

With the knowledge of $S_{\underline{\sigma}\,\underline{\sigma}'}$, we are now able to carry out signed
traces of the form
\eq{
\Tre O_+ - \Tro O_+ = \sum_{\underline{\sigma},\underline{\sigma}'}
S_{\underline{\sigma}\,\underline{\sigma}'} \Omega_{\underline{\sigma}\,\underline{\sigma}'}.
\label{str}}
Note that the terms in the sum of Eq.~\eref{str} completely factorize in the quadruplet index $k$. Furthermore, using the fact that $\Treo O_- = (\Treo O_+)^*$ and thus
$\Treo \rho^{T_2} = \rp \Treo O_+ + \ip \Treo O_+$, a simple
calculation leads to
\eq{
\fl
1-2\Tro \rho^{T_2} = \Tre \rho^{T_2} -\Tro \rho^{T_2} =
\prod_{k\in\mathrm{(I)}} \frac{1+|z_k|^2+2\ip z_k}{2}
\prod_{k\in\mathrm{(II)}} \frac{1+u_k v_k + u_k+v_k}{2}.
\label{str2}
}
Finally, we define the quantity
\eq{
\eo = \ln \max(1-2\Tro \rho^{T_2},1),
\label{eo}}
which clearly gives a lower bound for the logarithmic negativity, $\lneg \ge \eo$, with strict
equality if all the negative eigenvalues reside in the odd sector and their number
is equal to the dimension of that subspace. This is true for pure states and one expects it to be
valid for thermal states in a finite regime of temperatures above the ground state.
For general bipartite states, however, the dimension of the negative subspace can be much larger
\cite{johnston13, rana13}.

\begin{figure}[thb]
\center
\psfrag{E}[][][.8]{$\lneg, \eo$}
\includegraphics[width=0.49\columnwidth]{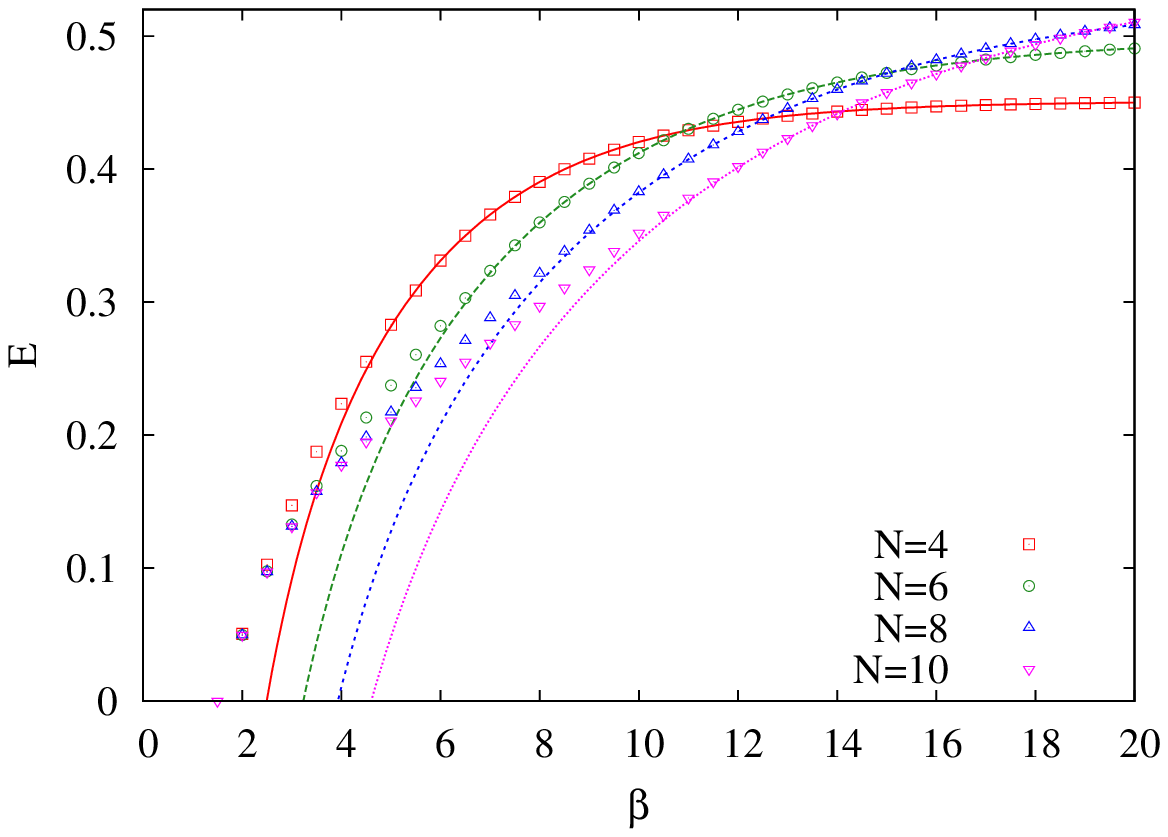}
\includegraphics[width=0.49\columnwidth]{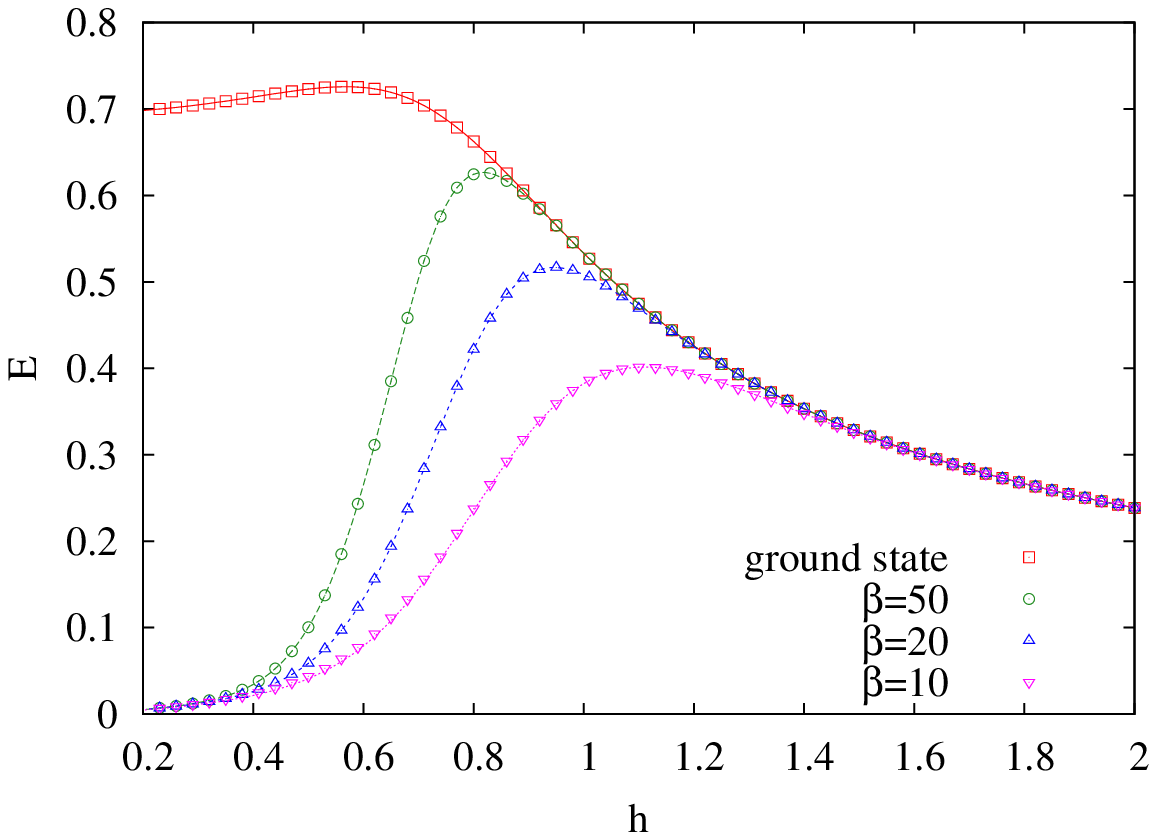}
\caption{Logarithmic negativity $\lneg$ (symbols) vs. $\eo$ (lines) between two halves
of a small Ising chain with Hamiltonian \eref{hti} in a Gibbs state.
Left: as a function of the inverse temperature $\beta$ with $h=1$ and various number of spins $N$.
The curves $\lneg$ and $\eo$ start to deviate around $\beta \approx N$.
Right: as a function of $h$ for various $\beta$ and $N=8$.
Minor deviations between $\lneg$ and $\eo$ are only visible for $\beta=10$.}
\label{fig:lno}
\end{figure}

To test the bound $\eo$ against the
exact value of the logarithmic negativity, we considered small TI chains with $N\le 10$
and obtained $\lneg$ via exact diagonalisation of $\rho^{T_2}$. This is shown on Fig.~\ref{fig:lno},
as a function of the temperature (left) as well as of the magnetic field (right).
We find that, for low enough temperatures, $\eo$ indeed exactly coincides with $\lneg$.
For larger temperatures, however, some of the negative eigenvalues in the odd sector
become positive or, vice versa, even eigenvectors could attain negative eigenvalues,
with both of these processes increasing the difference $\lneg-\eo$. Nevertheless,
for the small system sizes considered, it appears that $\eo$ gives a very good approximation
up to temperatures $T \approx 1/N$, above which it starts to deviate significantly.

%
%

\subsection{Ground states in a tripartite geometry}

So far we have only considered bipartite geometries. Another interesting setting
we are able to deal with is a pure state with the symmetric tripartite geometry, depicted on
Fig.~\ref{fig:subsys}(b). In this case, the reduced state $\rho_A$ after tracing out the sites
of $B$ is a mixed Gaussian state, associated to the reduced covariance matrix $\Gamma_A$,
with indices running over sites in $A$ \cite{LR09}.

The logarithmic negativity for the tripartite case can be obtained with CFT methods \cite{CCT12}.
For two intervals of the same size $\ell$ embedded in a system of length $L$ with periodic
boundary conditions, the calculation yields \cite{CCT13}
\eq{
\lneg(\ell,L) = \frac{c}{4}
\ln \left[ \frac{L}{\pi} \frac{\sin^2 \left(\frac{\pi \ell}{L} \right)}{
\sin \left(\frac{2\pi \ell}{L}\right)} \right] + \mathrm{const.} \, ,
\label{lnpbc}}
with the central charge $c$ and a non-universal constant. However, subtracting the value at
$\ell=L/4$ one obtains a universal scaling function
\eq{
\epsilon(z) = \lneg(\ell,L) -\lneg(L/4,L) =
\frac{1}{8}\ln \left[\tan (\pi z) \right],
\label{epbc}}
where $z=\ell/L$ and we have set $c=1/2$ corresponding to the TI chain.
The formula was tested using tensor network methods for the calculation of the
partial transpose, and a very good agreement was found \cite{CTT13}.

It is interesting to check the behaviour of the lower bound, defined in Eq. \eref{eo},
for the geometry at hand. In fact, since reflection symmetry is fulfilled, all the arguments
of the previous section, leading to Eq.~\eref{str2}, apply and $\eo$ is given by the same
formula in terms of the eigenvalues of $\Gamma_+$. However, there is an important
difference compared to the bipartite thermal case, which is apparent from the numerical
investigation of small systems. Namely, the number of negative eigenvalues of $\rho^{T_2}_A$
is always less then the dimension of the odd subspace and, moreover, some of the corresponding
eigenvectors are even. Thus, by tracing out the sites of $B$, the partial transpose
$\rho^{T_2}_A$ cannot be smoothly deformed from the pure-state case and, consequently,
one does not have a finite regime of parameters where the bound given by $\eo$ is tight.

\begin{figure}[htb]
\center
\psfrag{z=l/L}[][][.8]{$z=\ell/L$}
\psfrag{El}[][][.8]{$\eo(\ell,L) -\eo(L/4,L)$}
\psfrag{Eh}[][][.8]{$\eo(L/4,L)$}
\includegraphics[width=0.49\columnwidth]{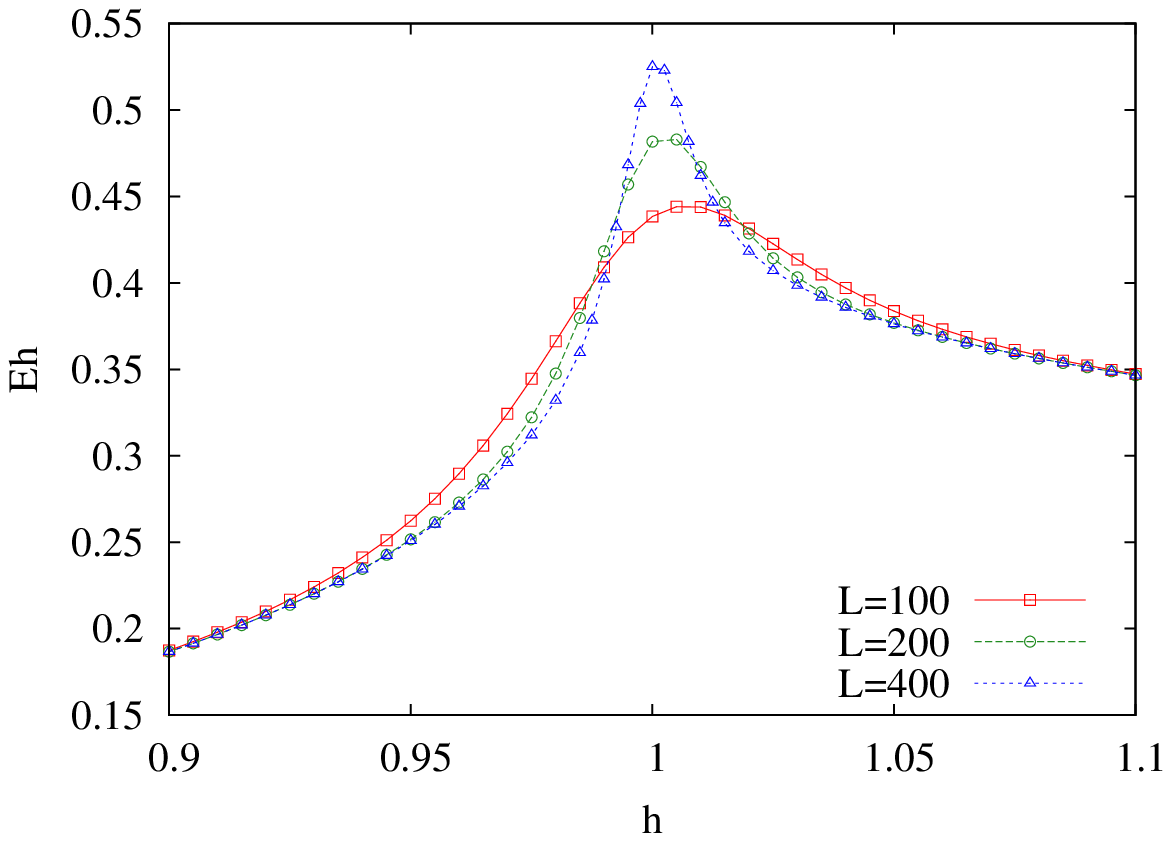}
\includegraphics[width=0.49\columnwidth]{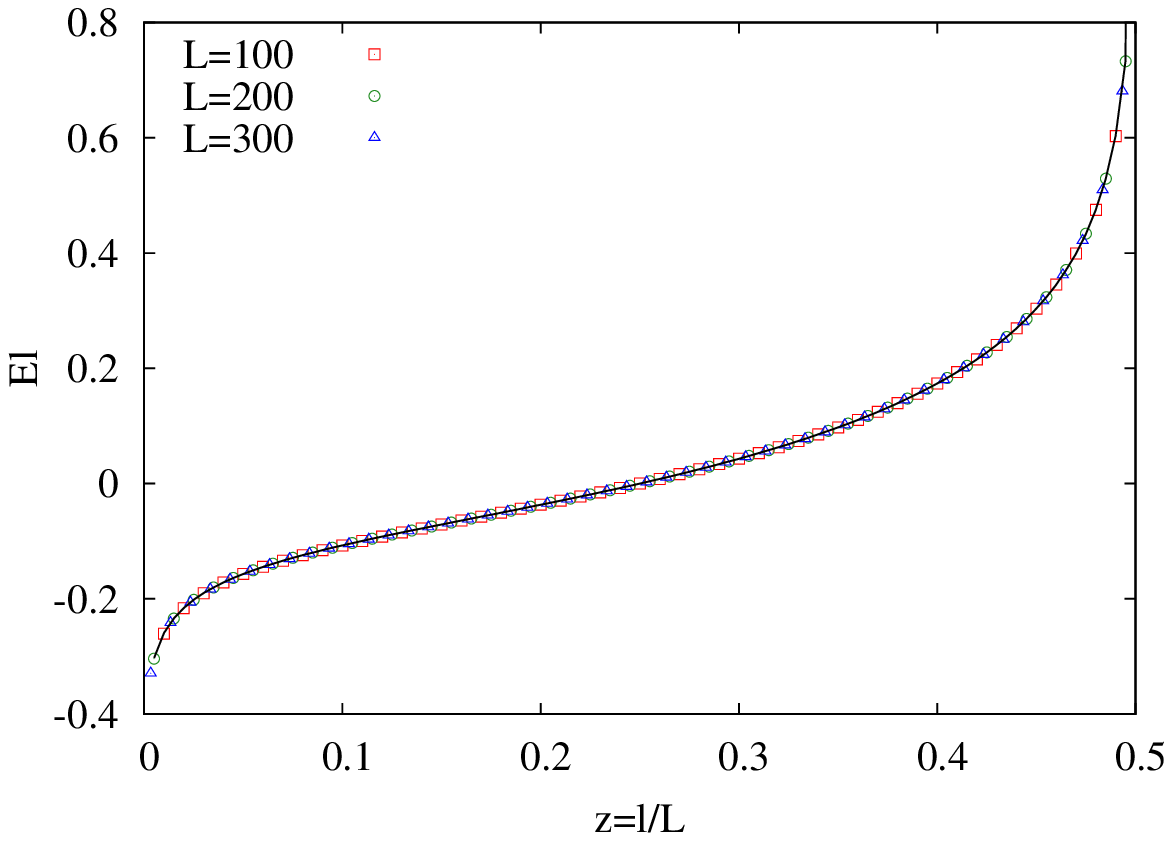}
\caption{The scaling of $\eo(\ell,L)$ for two adjacent intervals of equal length $\ell$
in the ground state of a periodic Ising chain with $L$ sites and magnetic field $h$.
Left: $\eo(L/4,L)$ as a function of the magnetic field $h$ for various $L$.
The logarithmic divergence around $h=1$ is clearly seen.
Right: $\eo(\ell,L)-\eo(L/4,L)$ at the critical point $h=1$ against the scaling variable $z$.
The solid line shows the scaling function in Eq. \eref{eopbc}.}
\label{fig:lnp}
\end{figure}

In spite of the above findings, $\eo$ shows a very interesting behaviour, which is
demonstrated on Fig.~\ref{fig:lnp}. First of all, it shows a clear signature of the phase
transition at $h=1$, which can be seen when plotting $\eo(L/4,L)$ against $h$
on the left of Fig.~\ref{fig:lnp}. Furthermore, defining the quantity $\epsilon_\mathrm{o}$
analogously to Eq. \eref{epbc}, one finds an excellent data collapse when plotted against
the variable $z=\ell/L$, see right of Fig. \ref{fig:lnp}.
The scaling function is found to be given by
\eq{
\epsilon_\mathrm{o}(z) = \eo(\ell,L) -\eo(L/4,L) =
\frac{1}{16}\ln \left[\frac{\tan (\pi z)}{2\cos^2(\pi z)} \right].
\label{eopbc}}

Although the functional form of $\epsilon_\mathrm{o}(z)$ was found by trial, one has
an excellent match with the data without any fitting parameters involved.
Interestingly, the prefactor of the logarithm is exactly the half of $\epsilon(z)$ in
Eq.~\eref{epbc}, however, the argument is modified as well.
We also performed a calculation directly in the thermodynamic limit $L\to\infty$,
and found $\eo(\ell) = 1/16 \ln \ell + \mathrm{const.}$, which is perfectly
consistent with the above findings. Furthermore, one could also consider the simple
fermionic hopping chain (or XX chain in spin language), defined by $B_{mn}=0$ and
$A_{mn}=\frac{1}{2}(\delta_{m,n-1} + \delta_{m,n+1})$. In complete analogy with the result for the
bipartite entanglement \cite{IJ08}, one finds $\eo^{XX}(2\ell)=2\eo^{TI}(\ell)$ for $h=1$,
and thus a doubled prefactor $1/8$ with respect to the critical TI chain.
Therefore, even though $\eo$ does not approximate $\lneg$ well, it shows exactly
the same universal behaviour, suggesting it as an entanglement indicator which is
extremely simple to calculate.

%
%

\section{Trace formulas\label{sec:tr}}

Our result for the partial transpose, Eq.~\eref{ptrho}, can be further tested by looking at
traces of integer powers of the partial transpose $\rho^{T_2}_A$ which can also be
carried out within CFT. Since the identity $\Tr (\rho^{T_2}_A)^2=\Tr \rho^2_A$ holds 
for any density matrix, the simplest nontrivial quantity to check is the trace of
the third power.
For the geometry of the previous section, one finds the CFT result \cite{CCT13}
\eq{
R_{3}(\ell,L) = 
\ln \Tr (\rho_A^{T_2})^3 = -\frac{1}{9}
\ln \left[ \frac{L^3}{\pi^3} \sin^2 \left(\frac{\pi \ell}{L} \right)
\sin \left(\frac{2\pi \ell}{L} \right) \right] + \mathrm{const.}
\label{R3cft}}
Similarly to Eq.~\eref{epbc}, a universal scaling function can be defined as \cite{CCT13}
\eq{
r_3(z) = R_3(\ell,L)-R_3(L/4,L)
= - \frac{1}{9} \ln \left[ 2\sin^2(\pi z) \sin(2\pi z) \right],
\label{r3cft}}
which was already tested numerically for the critical TI chain \cite{CTT13,Alba13}.
On the other hand, one could also consider two adjacent intervals of equal length $\ell$,
embedded in an infinite chain which is thermalized at inverse temperature $\beta$.
Applying a simple conformal transformation, the corresponding CFT formula follows as
\eq{
R_3(\ell,\beta)= - \frac{1}{9}
\ln \left[ \frac{\beta^3}{\pi^3} \sinh^2 \left(\frac{\pi \ell}{\beta} \right)
\sinh \left(\frac{2\pi \ell}{\beta} \right) \right] + \mathrm{const.}
\label{r3cft2}}

We now show how the above traces can be calculated with the covariance matrix formalism.
Expanding the third power of $\rho^{T_2}_A$ in Eq. \eref{ptrho} and taking the trace one arrives at
\eq{
\Tr (\rho_A^{T_2})^3 = -\frac{1}{2}\Tr (O^3_+) + \frac{3}{2} \Tr (O^2_+O_-),
}
where we have used that both of the traces on the right hand side are real.
In order to evaluate them, one has to invoke the determinant formulas for the
trace of products of Gaussian operators, which have already been considered
in different contexts \cite{FC10,BGZ14}. The main steps of this calculation are
summarized in \ref{app3}. In turn, one finds
\eq{
\Tr (\rho_A^{T_2})^3 = 
\mp \frac{1}{2} \sqrt{ \det \left( \frac{1+3\Gamma_+^2}{4}\right)} +
\frac{3}{2}\sqrt{ \det \left(\frac{1+\Gamma_+^2 + 2\Gamma_+ \Gamma_-}{4}\right)},
\label{r3det}}
where the sign of the first term depends on the spectrum of $\Gamma_+$, see Eq.~\eref{quad}.
Namely, the $+$ sign applies only if the quadruplet with purely imaginary
eigenvalues appears (see \ref{app3} for a more detailed discussion).
Note that similar sign ambiguities also appeared in \cite{FC10}. One should also remark,
that traces of higher powers can be handled in a very similar way, however, the
formulas become rather lengthy.

\begin{figure}[thb]
\center
\psfrag{z=l/L}[][][.8]{$z=\ell/L$}
\psfrag{l}[][][.8]{$\ell$}
\psfrag{R3}[][][.8]{$R_3(\ell,L) -R_3(L/4,L)$}
\psfrag{-R3}[][][.8]{$-R_3(\ell,\beta)$}
\psfrag{CFTb}[][][.55]{$\frac{\beta^3}{\pi^3} \sinh^2 \left(\frac{\pi \ell}{\beta} \right)
\sinh \left(\frac{2\pi \ell}{\beta} \right)$}
\includegraphics[width=0.49\columnwidth]{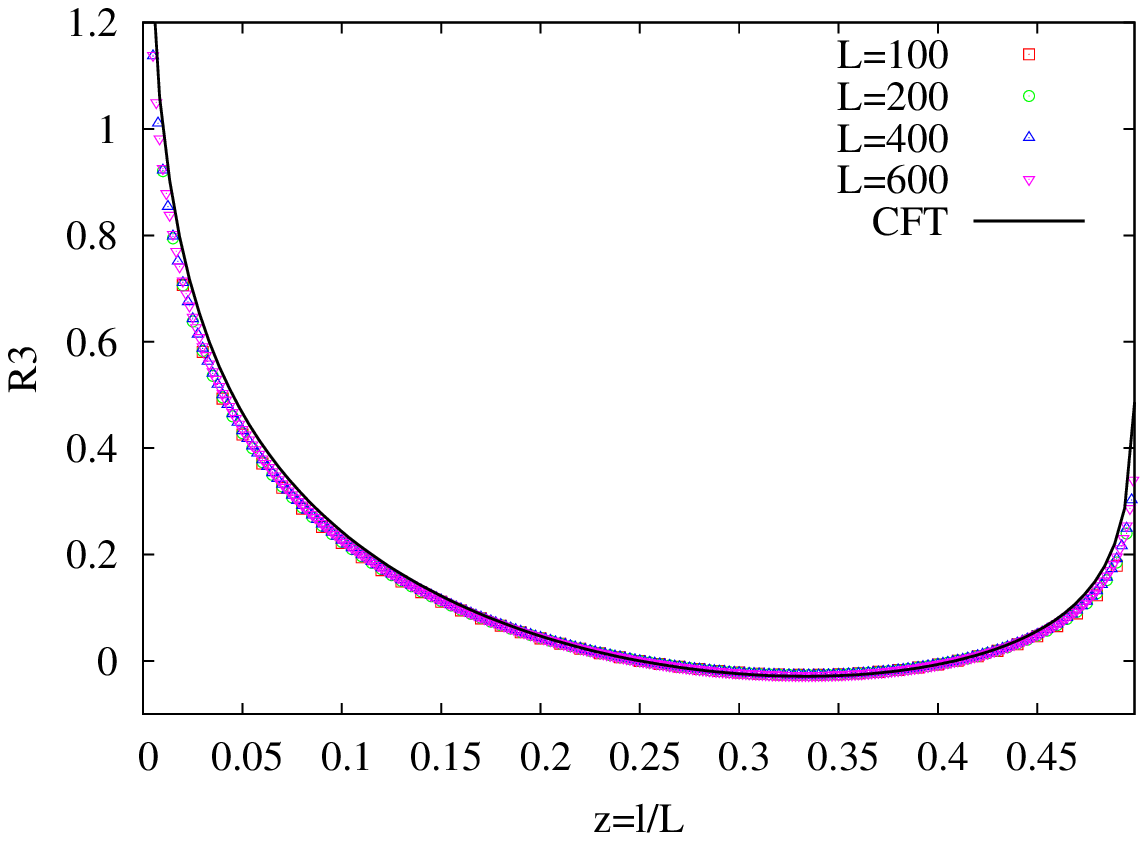}
\includegraphics[width=0.49\columnwidth]{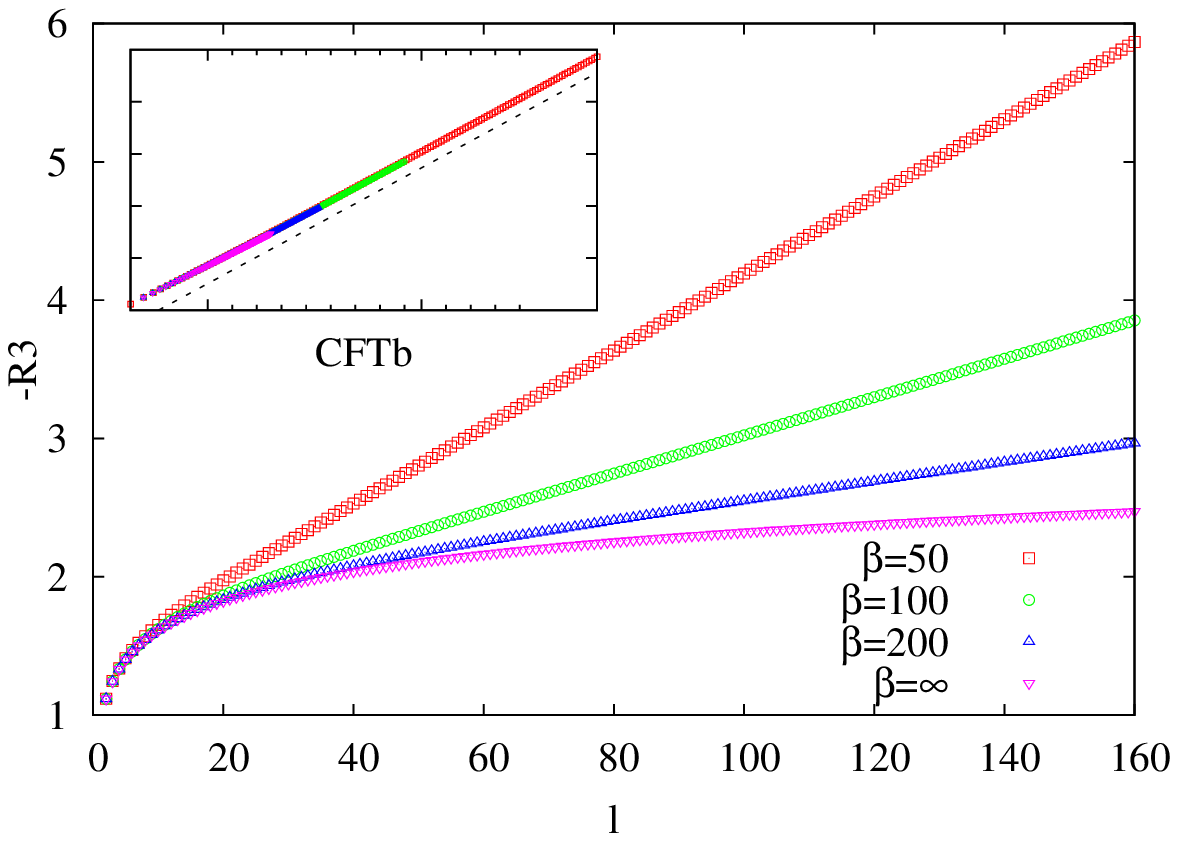}
\caption{Left: $R_3(\ell,L)-R_3(L/4,L)$ as function of $z=\ell/L$ for two adjacent
intervals of length $\ell$ in the ground state of the critical TI chain of size $L$.
The solid line shows the CFT formula \eref{r3cft}.
Right: $R_3(\ell,\beta)$ for adjacent intervals of size $\ell$ in a thermal state of the
infinite chain ($h=1$), with inverse temperature $\beta$.
The inset shows the rescaled data compared to the CFT prediction \eref{r3cft2}
on a horizontal log-scale.}
\label{fig:r3}
\end{figure}

The trace formula \eref{r3det} can now be compared to the CFT predictions in \eref{r3cft}
and \eref{r3cft2} by inserting the corresponding covariance matrices $\Gamma_\pm$
and evaluating the determinants. This is shown in Fig.~\ref{fig:r3} for the ground (left)
and thermal states (right), respectively. The perfect agreement of the curves
provides a highly nontrivial check of the CFT results.

%
%

\section{Discussion
\label{sec:disc}}

In conclusion, we have shown that the partial transpose of fermionic Gaussian states
can be written as a linear combination of only two Gaussian operators, uniquely determined
by associated covariance matrices $\Gamma_\pm$. In the presence of reflection symmetry,
this particular form of the partial transpose allows us to carry out  traces
over the even/odd subspaces which, in turn, can be used to construct a lower bound to
the logarithmic negativity. Furthermore, the trace of 
any integer power of $\rho^{T_2}_A$ can, in principle, be calculated as a sum of determinants,
each of linear size $2|A|$.

There are several open questions left for future research. We did not consider
in detail entanglement detection questions, e.g., providing temperature bounds 
for separability of fermion or spin systems in thermal equilibrium. It would be instructive to 
compare such results obtained from the negativity lower bound $\eo$ with 
earlier studies \cite{Toth05, Wu05,GT09}.

Another natural extension of our work would be to consider non-adjacent intervals. 
For spin chains, however, it was shown that the spin RDM itself is already a linear
combination of four fermionic RDMs \cite{FC10}. Although our construction for the
partial transpose could be carried over, it would further double the number of terms
in the linear combination. Thus, the calculation of the traces for such a case is still
realisable, but presumably more tedious.

It would also be interesting to see whether the lower bound $\eo$ could be attainable
within the framework of CFT.
This could lead to an analytical understanding of the scaling function for the critical tripartite
case in Eq.~\eref{eopbc} and could shed light to the origin of the prefactor.
In fact, one is tempted to guess that this is equal to one-half of the corresponding 
prefactor of the logarithmic negativity in a general CFT. From the free-fermion point of view,
our analysis clearly suggests that certain asymptotic relations between $\eo$ and $\lneg$
could hold in general. Finding a rigorous form of this relation would allow for a numerically
feasible estimation of the entanglement negativity for fermionic systems.

Finally, we point out that 
some specific classes of mixed Gaussian states exist which
allow for an exact calculation of the logarithmic negativity using the methods introduced here.
These are the states for which the relation $\left[\Gamma_+,\Gamma_-\right]=0$ is satisfied, an example being the isotropic Gaussian states \cite{BR04}, for which the covariance matrix satisfies $\Gamma^2=\lambda^2 \identity$ with some $0\le\lambda<1$.
The situation is then analogous to the pure-state case and the corresponding calculation can
be generalized, which we leave for future studies.

\ack{
We would like to thank Leonardo Banchi and Fernando Brand\~{a}o for useful discussions,
and an anonymous referee for pointing out Ref. \cite{BR04}.
The work of V.E. was supported by OTKA Grant No. NK100296; and
Z.Z. acknowledges funding by the British Engineering and Physical Sciences Research Council	(EPSRC).}

\appendix

\section{The partial transpose of a 2-site RDM
\label{app2}}

It is instructive to check how the method introduced in Section \ref{sec:pt}  works
for the simplest case of two consecutive sites.
The canonical form of the Gaussian RDM is given by
\eq{
\rho = \prod_{k=1,2} \frac{\identity + i\nu_k b_{2k-1}b_{2k}}{2},
\label{gs2}}
where $\nu_k \in \left[0,1\right]$ and  the Majorana operators $b_j$ are related to $a_j$ through an orthogonal
transformation.
For simplicity, we will consider only covariance matrices of the
form of Eq.~\eref{gamma}, and assume also reflection symmetry. These states are parametrized
by a single  angle $\theta$ beside the covariance matrix eigenvalues $\nu_k$.

Using the Jordan-Wigner representation of $a_j$ and
working in the usual spin basis, the most general form of the partial transpose is
\eq{\fl
\rho^{T_2}= 
\frac{\identity}{4} +
\frac{\nu_1\nu_2}{4} \sigma^z_1\sigma^z_2+
\frac{\nu_1+\nu_2}{4} 
\scalebox{0.7}{
$\left(\begin{array}{cccc}
\cos 2\theta &  &  & 0\\
  & 0 & \sin 2\theta & \\
  & \sin 2\theta & 0 &  \\
0 &  &  & -\cos 2\theta
\end{array} \right)$} +
\frac{\nu_1-\nu_2}{4} 
\scalebox{0.7}{
$\left(\begin{array}{cccc}
0 &  &  & 1\\
  & 0 & 0 & \\
  & 0 & 0 &  \\
1 &  &  & 0
\end{array} \right)$}.
}
Note that, besides the diagonals, all matrix elements vanish and are thus not shown.
It is straightforward to obtain the four eigenvalues
\eq{
\eqalign{
\lambda_{1,2} = \left[1+\nu_1\nu_2 \pm
\sqrt{(\nu_1+\nu_2)^2 \cos^2 2\theta + (\nu_1-\nu_2)^2}\right]/4, \\
\lambda_{3,4} = \left[1-\nu_1\nu_2 \pm (\nu_1+\nu_2) \sin 2\theta \right]/4.}
\label{lam2sp}}
Using the parity operator $P_{2}=\sigma^z_2$, one can also construct
$\rho^{T_2}_{\pm} = (\rho^{T_2} \pm P_{2} \rho^{T_2} P_{2})/2$ with matrix elements
\begin{eqnarray}
\rho_+^{T_2} &= \frac{\identity}{4} +
\frac{\nu_1\nu_2}{4} \sigma^z_1\sigma^z_2+
\frac{\nu_1+\nu_2}{4} 
\scalebox{0.7}{$
\left(\begin{array}{cccc}
\cos 2\theta &  &  & 0\\
  & 0 & 0 & \\
  & 0 & 0 &  \\
0 &  &  & -\cos 2\theta
\end{array} \right)$}, \\
\rho_-^{T_2} &=
\frac{\nu_1+\nu_2}{4} 
\scalebox{0.7}{$
\left(\begin{array}{cccc}
0 &  &  & 0\\
  & 0 & \sin 2\theta & \\
  & \sin 2\theta & 0 &  \\
0 &  &  & 0
\end{array} \right)$} +
\frac{\nu_1-\nu_2}{4}
\scalebox{0.7}{$
\left(\begin{array}{cccc}
0 &  &  & 1\\
  & 0 & 0 & \\
  & 0 & 0 &  \\
1 &  &  & 0
\end{array} \right)$}.
\label{r-t2}
\end{eqnarray}
The eigenvalues of the operator $\rho_+^{T_2} + i \rho_-^{T_2}$ then read
\eq{
\eqalign{
\Omega_{1,2} = \left[1+\nu_1\nu_2 \pm
\sqrt{(\nu_1+\nu_2)^2 \cos^2 2\theta - (\nu_1-\nu_2)^2}\right]/4, \\
\Omega_{3,4} = \left[1-\nu_1\nu_2 \pm i (\nu_1+\nu_2) \sin 2\theta \right]/4.}
\label{omi}}
Note that we have $\lambda_{3,4} = \rp \Omega_{3,4} + \ip \Omega_{3,4}$ which
simply follows from the fact that $\rho^{T_2}_{+}$ and $\rho^{T_2}_{-}$ commute on the corresponding subspace,
including the odd eigenvector.
Unfortunately, this property does not generalize to symmetrically bipartitioned intervals with more than two spins.

We will now show that the Gaussian operator $O_+$ with covariance matrix $\Gamma_+$ has indeed
eigenvalues given by Eq.~\eref{omi}.
The covariance matrix $\Gamma$ for the Gaussian state \eref{gs2} and the associated $\Gamma_+$
have the form
\eq{\fl
\Gamma = i\left(
\begin{array}{cccc}
0 & c & 0 & s_- \\
-c & 0 & s_+ & 0\\
0 & -s_+ & 0 & c \\
-s_- & 0 & -c & 0
\end{array}
\right), \qquad
\Gamma_{+} =  i\left(
\begin{array}{cccc}
0 & c & 0 & is_- \\
-c & 0 & is_+ & 0\\
0 & -is_+ & 0 & -c \\
-is_- & 0 & c & 0
\end{array}
\right),
}
with the shorthand notation
\eq{
c = \frac{\nu_1+\nu_2}{2} \cos 2\theta, \qquad
s_{\pm} = \frac{\nu_1+\nu_2}{2} \sin 2\theta \pm \frac{\nu_1-\nu_2}{2}.
\label{cs}}
The four eigenvalues $\pm \nu^{\pm}$ of $\Gamma_+$ can be computed with
\eq{
\nu^{\pm} = 
\sqrt{c^2-\left(\frac{s_+-s_-}{2}\right)^2} \pm i \frac{s_+ + s_-}{2}.
\label{nupm2}}
If the operator $O_+$ is Gaussian, its eigenvalues must have the form
\eq{
\eqalign{
\Omega_{++} = \frac{1+\nu_+}{2}\frac{1+\nu_-}{2}, \quad
\Omega_{--} = \frac{1-\nu_+}{2}\frac{1-\nu_-}{2}, \\
\Omega_{+-} = \frac{1+\nu_+}{2}\frac{1-\nu_-}{2}, \quad
\Omega_{-+} = \frac{1-\nu_+}{2}\frac{1+\nu_-}{2}.}
\label{ompm}}
Substituting \eref{cs} and \eref{nupm2} into \eref{ompm}, we indeed recover
the values in \eref{omi}.

Finally, let us shortly discuss the non-Gaussian character of $\rho^{T_2}$. 
comparing the expectation values $\Tr(\rho^{T_2} a_m a_n)$, where $m, n{=}1,{\ldots}, 4$,
with $\Tr(\rho^{T_2} a_1 a_2 a_3 a_4)$, one observes that the Wick expansion, 
Eq.~\eref{wick}, does not hold, unless $c^2 = \nu_1 \nu_2$. This remains true whatever 
basis we choose for the partial transpose. Thus, the partial transpose of a 
fermionic Gaussian state is usually not a Gaussian operator.

\section{Eigenvalues of $\Gamma_{\pm}$ for pure states
\label{app1}}

Here we consider the eigenvalue problem of the modified covariance matrices $\Gamma_\pm$,
that are associated to a pure-state covariance matrix $\Gamma$ of the form \eref{gamma}.
The RDMs for subsystems $A_1$ and $A_2$ are determined via the reduced covariance matrices
$\Gamma^{11}$ and $\Gamma^{22}$. Since they split into two
submatrices, one could equivalently consider the squared eigenvalue problem of the matrices
\eq{\fl
G^{\alpha\alpha}_{mn} = \sum_{l\in\alpha} g_{ml} g_{nl} =
\sum_{p,q} M^{\alpha}_{pq}\psi^{\alpha}_{p}(m) \psi^{\alpha}_q(n), \qquad
\psi^{\alpha}_{q}(m) = 
\cases{
\psi_q(m) & $m \in A_\alpha$, \\ 0 &  $m \notin A_\alpha$,
}
\label{gam}}
with nonzero matrix elements only within $m,n \in A_\alpha$, $\alpha =1,2$.
The overlap matrices $M^{\alpha}$ have matrix elements
\eq{
M^{\alpha}_{pq} = \sum_{l} \phi^{\alpha}_p(l) \phi^{\alpha}_q(l), \qquad
\phi^{\alpha}_q(l) =
\cases{
\phi_q(l) & $l \in A_\alpha$, \\ 0 & $l \notin A_\alpha$.
}
}
Note that both $G^{11}$ and $G^{22}$ have the same eigenvalues
$\mu^2_k \le 1$ with $k=1,\dots,\min(|A_1|,|A_2|)$, whereas $\mu^2_k=1$ for the
remaining eigenvalues of the larger matrix. We also introduce the block-diagonal matrix
\eq{
G = \twomat{G^{11}}{0}{0}{G^{22}},
}
with all the nontrivial eigenvalues being doubly degenerate.

The matrix elements of the covariance matrices $\Gamma_{\pm}$ in Eq. \eref{gammapm}
are determined through
\eq{
g^{\pm}_{mn} = \sum_{q} \psi^{\pm}_q(m) \phi^{\pm}_q(n),
\label{gpm}}
with the vectors
\eq{
\phi^{\pm}_q(l) = \phi^{1}_q(l) \pm i \phi^{2}_q(l), \qquad
\psi^{\pm}_q(l) = \psi^{1}_q(l) \pm i \psi^{2}_q(l) \, .
\label{phipm}}
Thus the spectrum of $\Gamma_{\pm}$ follows from the eigenvalues of the squared matrix
\eq{
G^{\pm}_{mn} = \sum_{p,q} (M^{1}_{pq}-M^{2}_{pq})\psi^{\pm}_p(m) \psi^{\pm}_q(n) \, .
\label{Gpm}}
Inserting \eref{phipm} and using the completeness property $M^{1}_{pq} + M^{2}_{pq} = \delta_{pq}$,
the matrices $G^{\pm}$ have the block form
\eq{
G^{\pm} = \twomat{2G^{11} - \identity}{\pm 2i F}{\pm 2i F^T}{2G^{22}-\identity},
}
with
\eq{
F_{mn} = \sum_{p,q} M^{1}_{pq} \psi^{1}_p(m) \psi^{2}_q(n).
\label{f}}
It is easy to check that $F$ satisfies
\eq{
F F^{T} = G^{11} (\identity - G^{11}), \qquad
F^T F = G^{22} (\identity - G^{22}),
}
and thus the following matrix identity holds
\eq{
(G^{\pm} - 2G + \identity)^2 = -4G(\identity-G).
}
Rewriting in terms of the eigenvalues $(\nu^{\pm}_k)^2$ and $\mu^2_k$
of $G^{\pm}$ and $G$, respectively, one arrives at
\eq{
(\nu^{\pm}_k)^2 = 2 \mu^2_k - 1 \pm 2i\mu_k \sqrt{1-\mu^2_k} =
\left(\mu_k \pm i \sqrt{1 - \mu_k^2}\right)^2.
\label{munu}}

\section{Determinant formulas
\label{app3}}

Let us consider the Gaussian operators $O_\pm$ corresponding to the generalized covariance
matrices $\Gamma_\pm  = \tanh(W_\pm/2)$. With $a$ denoting the vector of Majorana operators,
we introduce
\eq{
O\left( W_\pm\right) = \exp\left(\frac{a W_\pm a}{4}\right), \qquad
Z(W_\pm) = \Tr \exp\left(\frac{a W_\pm a}{4}\right),
}
and thus $O_\pm = O\left(W_\pm\right)/Z(W_\pm)$
Our aim is to calculate various traces of the form $\Tr (O^m_+O^n_-)$ with some integers $m$ and $n$.
Following the lines of Ref.~\cite{FC10}, we first introduce the notation
\eq{
\left\{W_1, W_2\right\} = \Tr \left[ O\left( W_1\right) O\left(W_2\right) \right].
}
We also note the simple fact that $O^m\left(W_\pm\right) = O\left(m W_\pm\right)$.
Hence, the traces we consider can be written in the form
\eq{
\Tr(O^m_+) = \frac{Z(mW_+)}{Z^m(W_+)}, \qquad
\Tr(O^m_+ O^n_-) = \frac{\left\{ mW_+,nW_- \right\}}{Z^m(W_+)Z^n(W_-)}.
\label{trmn}}
They can be evaluated in terms of determinant formulas \cite{FC10}
\eq{\fl
Z(W) = (\pm)\sqrt{\det \left(2 \cosh\frac{W}{2}\right)}, \qquad
\left\{W_1, W_2\right\} = (\pm)\sqrt{\det(1+\ee^{W_1}\ee^{W_2})},
\label{trdet}}
where the $\pm$ in parentheses symbolise the eventual sign ambiguity.
The square root (and hence the sign ambiguity) indicates that
the pairs of eigenvalues of the skew-symmetric matrices
must be taken into account with halved degeneracy \cite{FC10}. Note that, for
Gaussian states commuting with the particle number operator (i.e., when
the exponent can be written with a Hermitian matrix in terms of the fermion
operators instead of Majoranas), similar trace formulas apply without square
roots and sign ambiguity \cite{Klich02}.

In Section~\ref{sec:tr} we need the traces of operators $O^3_+$ and $O^2_+O_-$,
respectively. Applying Eqs.~\eref{trmn} and \eref{trdet}, using hyperbolic identities
for multiple arguments, one observes that the formulas can be expressed solely in terms of
$\Gamma_\pm$ with the result
\eq{\fl
\Tr(O^3_+) = \pm \sqrt{ \det \left( \frac{1+3\Gamma_+^2}{4}\right)} \, , \qquad
\Tr(O^2_+O_-) = \sqrt{ \det \left(\frac{1+\Gamma_+^2 + 2\Gamma_+ \Gamma_-}{4}\right)} \, .
\label{tr12}}
The sign ambiguity can be fixed by comparing to exact calculations of the traces.
We find that the negative sign in the first trace of Eq.~\eref{tr12} is needed only in case
$\Gamma_+$ contains a quadruplet of purely imaginary eigenvalues. For the Ising chain,
this can happen only in the symmetry-broken phase, $h<1$. The numerics for small chains
shows that, gradually decreasing the value of $h$, the appearance of this quadruplet
exactly coincides with the vanishing of the first determinant. Interestingly, the second determinant
in Eq.~\eref{tr12} always remains positive and thus no sign ambiguity appears there. 

\section*{References}

\bibliographystyle{iopart-num.bst}

\bibliography{transpose}

\end{document}